\documentclass[aps,pra,showpacs,preprint]{revtex4}
\usepackage{graphicx}
\usepackage{amsmath,amssymb,amsfonts}
\usepackage{dcolumn}
\usepackage[english]{babel}

\usepackage{epsfig}

\begin{document}

\title{The parabolic Sturmian-function basis representation of the
six-dimensional Coulomb Green's function}
\author{ S. A. Zaytsev}
\email[E-mail: ]{zaytsev@fizika.khstu.ru} \affiliation{Pacific
National University, Khabarovsk, 680035, Russia}

\begin{abstract}
The square integrable basis set representation of the resolvent of
the asymptotic three-body Coulomb wave operator in parabolic
coordinates is obtained. The resulting six-dimensional Green's
function matrix is expressed as a convolution integral over
separation constants.
\end{abstract}
\pacs{03.65.Nk} \maketitle

\section{Introduction}
It is well known that the Schr\"{o}dinger equation for a three-body
Coulomb system is asymptotically separable in terms of parabolic
coordinates \cite{Klar}
\begin{equation}\label{PC}
    \xi_j = r_{ls}+\hat{\bf k}_{ls}\cdot {\bf r}_{ls}, \quad
    \eta_j = r_{ls}-\hat{\bf k}_{ls}\cdot {\bf r}_{ls},
\end{equation}
where ${\bf r}_{ls}$ and ${\bf k}_{ls}$ are the relative coordinate
and momentum vectors between the particles $l$ and $s$. Here $j$,
$l$, $s$ is a cyclic permutation of 1, 2, 3. The long-ranged
six-dimensional operator, which provides a three-body continuum wave
function, 3C-function \cite{C31,C32}, that satisfies exact
asymptotic boundary conditions for Coulomb systems in the region
where the distances between the particles are large, reads as the
sum of two-dimensional operators \cite{Klar}:
\begin{equation}\label{AO}
    \sum_{j=1}^3 \frac{1}{\mu_{ls}(\xi_j+\eta_j)}
    \left[ \hat{h}_{\xi_j}+\hat{h}_{\eta_j}+2k_{ls}t_{ls}\right],
\end{equation}
where $t_{ls}=\frac{Z_l Z_s \mu_{ls}}{k_{ls}}$, $\mu_{ls}=\frac{m_l
m_s}{m_l+m_s}$. Here the one-dimensional operators $\hat{h}_{\xi_j}$
and $\hat{h}_{\eta_j}$ are defined by
\begin{equation}\label{OO}
\begin{array}{ccc}
  \hat{h}_{\xi_j} &=& -2\left(\frac{\partial}{\partial\xi_j}\xi_j
    \frac{\partial}{\partial\xi_j}+ik_{ls}\xi_j
    \frac{\partial}{\partial\xi_j}\right),\\
  \hat{h}_{\eta_j} &=& -2\left(\frac{\partial}{\partial\eta_j}\eta_j
    \frac{\partial}{\partial\eta_j}-ik_{ls}\eta_j
    \frac{\partial}{\partial\eta_j}\right).
\end{array}
\end{equation}
In the previous paper \cite{previous} we introduce the
six-dimensional operator $\mathfrak{h}$ which is obtained by
multiplying (\ref{AO}) on the left by
$\prod_{j=1}^3\mu_{ls}(\xi_j+\eta_j)$:
\begin{equation}\label{h6}
\hat{\mathfrak{h}}=\mu_{13}(\xi_2+\eta_2)\,\mu_{12}(\xi_3+\eta_3)\hat{\mathfrak{h}}_1+
\mu_{23}(\xi_1+\eta_1)\,\mu_{12}(\xi_3+\eta_3)\hat{\mathfrak{h}}_2+
\mu_{23}(\xi_1+\eta_1)\,\mu_{13}(\xi_2+\eta_2)\hat{\mathfrak{h}}_3,
\end{equation}
where
\begin{equation}\label{h2}
\hat{\mathfrak{h}}_j=\hat{h}_{\xi_j}+\hat{h}_{\eta_j}+2k_{ls}t_{ls}.
\end{equation}
The resolvent of the operator (\ref{h6}) can be used in the
corresponding Lippmann-Schwinger equation for the three-body Coulomb
wave function.

It has been suggested \cite{previous} to treat the operator
(\ref{h6}) within the context of $L^2$ parabolic Sturmian basis set
\cite{Ojha1}
\begin{equation}\label{S6}
\left|\mathcal{N}\right>=\prod_{j=1}^3\phi_{n_j\,m_j}(\xi_j,\,\eta_j),
\end{equation}
\begin{equation}\label{S2}
  \phi_{n_j\,m_j}(\xi_j,\,\eta_j)=\psi_{n_j}(\xi_j)\,\psi_{m_j}(\eta_j),
\end{equation}
\begin{equation}\label{S1}
  \psi_n(x)=\sqrt{2b} e^{-bx}L_{n}(2bx),
\end{equation}
where $b$ is the scaling parameter. In particular, a matrix
representation ${\bf G}_j$ of the resolvent for the two-dimensional
operator
\begin{equation}\label{h2Q}
\hat{\mathfrak{h}}_j+\mu_{ls}C_j(\xi_j+\eta_j)
\end{equation}
has been obtained which is formally the matrix inverse to the
infinite matrix $\left[{\bf h}_j+\mu_{ls}C_j{\bf Q}_j\right]$ of the
operator (\ref{h2Q}):
\begin{equation}\label{mh2Q}
    \left[{\bf h}_j+\mu_{ls}C_j{\bf Q}_j\right]
    {\bf G}_j(t_{ls};\; \mu_{ls}C_j)={\bf I}_j.
\end{equation}
Here
\begin{equation}\label{mh2}
    {\bf h}_j={\bf h}_{\xi_j}\otimes{\bf I}_{\eta_j}+{\bf I}_{\xi_j}\otimes{\bf h}_{\eta_j}
    +2k_{ls}t_{ls}{\bf I}_j
\end{equation}
is the matrix of the operator $\hat{\mathfrak{h}}_j$ (\ref{h2}) in
the basis (\ref{S2}), ${\bf I}_{\xi_j}$, ${\bf I}_{\eta_j}$ and
${\bf I}_j={\bf I}_{\xi_j}\otimes{\bf I}_{\eta_j}$ are the unit
matrices. ${\bf Q}_j={\bf Q}_{\xi_j}\otimes{\bf I}_{\eta_j}+{\bf
I}_{\xi_j}\otimes{\bf Q}_{\eta_j}$, where ${\bf Q}_{\xi_j}$ and
${\bf Q}_{\eta_j}$ are the matrices of $\xi_j$ and $\eta_j$ in basis
(\ref{S1}), respectively.

In the previous paper \cite{previous} it has been suggested that the
matrices ${\bf G}_j$ of the two-dimensional Green's functions be
used to construct the matrix $\underline{\mathfrak{G}}$ which is
inverse to the operator (\ref{h6}) matrix
\begin{equation}\label{mh6}
    \underline{\mathfrak{h}}=\mu_{13}\mu_{12}{\bf h}_1\otimes{\bf Q}_2\otimes{\bf
    Q}_3+\mu_{23}\mu_{12}{\bf Q}_1\otimes{\bf h}_2\otimes{\bf
    Q}_3+\mu_{23}\mu_{13}{\bf Q}_1\otimes{\bf
    Q}_2\otimes{\bf h}_3.
\end{equation}
Namely, we proposed to express the six-dimensional Green's function
matrix $\underline{\mathfrak{G}}$ as the convolution integral
\begin{equation}\label{G6}
\underline{\mathfrak{G}}= \aleph \int\int dC_1 dC_2\,{\bf
G}_1(t_{23};\; \mu_{23}C_1)\otimes
    {\bf G}_2(t_{13};\; \mu_{13}C_2)\otimes{\bf G}_3(t_{12};\;
    -\mu_{12}(C_1+C_2)),
\end{equation}
where $\aleph$ is a normalizing factor. Thus, our problem is now to
determine the pathes of integration over the separation constants
$C_1$ and $C_2$ in (\ref{G6}) and to find the corresponding
normalizing factor $\aleph$ such that the condition
\begin{equation}\label{hGE}
    \underline{\mathfrak{h}}\underline{\mathfrak{G}}=
    {\bf I}_1\otimes{\bf I}_2\otimes{\bf I}_3
\end{equation}
holds. For this purpose consider the product
$\underline{\mathfrak{h}}\;\underline{\mathfrak{G}}$. From the
relation (\ref{mh2Q}) we obtain
\begin{equation}\label{hG2}
 \begin{array}{c}
\underline{\mathfrak{h}}\;\underline{\mathfrak{G}}= \aleph \left\{
\int\int dC_1 dC_2\left[{\bf I}_1-\mu_{23}C_1{\bf Q}_1{\bf
G}_1(t_{23};\; \mu_{23}C_1)\right]\otimes\mu_{13}{\bf Q}_2{\bf
G}_2(t_{13};\;
\mu_{13}C_2) \right.\\[3mm]\otimes\mu_{12}{\bf Q}_3{\bf G}_3(t_{12};\;
    -\mu_{12}(C_1+C_2))\\[3mm]
\int\int dC_1 dC_2\,\mu_{23}{\bf Q}_1{\bf G}_1(t_{23};\;
\mu_{23}C_1)\otimes\left[{\bf I}_2-\mu_{13}C_2{\bf Q}_2{\bf
G}_2(t_{13};\; \mu_{13}C_2)\right]\\[3mm]
\otimes\mu_{12}{\bf Q}_3{\bf G}_3(t_{12};\;-\mu_{12}(C_1+C_2))\\[3mm]
\int\int dC_1 dC_2\,\mu_{23}{\bf Q}_1{\bf G}_1(t_{23};\;
\mu_{23}C_1)\otimes\mu_{13}{\bf Q}_2{\bf
G}_2(t_{13};\; \mu_{13}C_2)\\[3mm]
\left. \otimes\left[{\bf I}_3+\mu_{12}(C_1+C_2){\bf Q}_3
{\bf G}_3(t_{12};\;-\mu_{12}(C_1+C_2))\right] \right\},\\
 \end{array}
\end{equation}
and hence
\begin{equation}\label{hG3}
 \begin{array}{c}
\underline{\mathfrak{h}}\;\underline{\mathfrak{G}}= \aleph
\left\{\int\int dC_1 dC_2\left[{\bf I}_1\otimes\mu_{13}{\bf Q}_2{\bf
G}_2(t_{13};\; \mu_{13}C_2)\otimes\mu_{12}{\bf Q}_3{\bf
G}_3(t_{12};\;
    -\mu_{12}(C_1+C_2))\right. \right.\\[3mm]
\left. +\mu_{23}{\bf Q}_1{\bf G}_1(t_{23};\; \mu_{23}C_1)\otimes{\bf
I}_2\otimes\mu_{12}{\bf Q}_3 {\bf G}_3(t_{12};\;-\mu_{12}(C_1+C_2))\right]\\[3mm]
\left.+\left[\mu_{23}{\bf Q}_1\int dC_1\,{\bf G}_1(t_{23};\;
\mu_{23}C_1)\right]\otimes\left[\mu_{13}{\bf Q}_2\int dC_2\,{\bf
G}_2(t_{13};\; \mu_{13}C_2) \right]\otimes{\bf I}_3 \right\}.
 \end{array}
\end{equation}
As a first step towards our goal we consider the integrals
\begin{equation}\label{Int}
 \int dC_j\,{\bf G}_j(t_{ls};\; \mu_{ls}C_j)
\end{equation}
inside the figure brackets on the right-hand side of (\ref{hG3}).

In Sec.~II completeness of the eigenfunctions of the two-dimensional
operator (\ref{h2Q}) is considered. In particular, an integral
representation of the matrix ${\bf A}$ which is inverse to the
infinite matrix ${\bf Q}$ of the operator $(\xi+\eta)$ in the basis
(\ref{S2}) is obtained. In Sec.~{III} it is demonstrated that the
integral (\ref{Int}) taken along an appropriate contour is
proportional to the matrix ${\bf A}$ obtained in the previous
section. Finally, Sec.~IV presents a convolution integral
representation of the six-dimensional Coulomb Green' function
matrix.

\section{Completeness relations}

\subsection{The continuous spectrum}
Of particular interest are the regular solutions
\begin{equation}\label{fuv} f(\gamma, \,
\tau, \,\xi, \, \eta)=u(\gamma, \, \tau, \, \xi)\,v(\gamma, \, \tau,
\, \eta)
\end{equation}
of the system
\begin{equation}\label{SYS1}
  \left[ \hat{h}_{\xi}+2kt+\mu\,C\xi\right] u(\gamma, \, \tau, \, \xi)=
  0,
\end{equation}
\begin{equation}\label{SYS2}
  \left[ \hat{h}_{\eta}+2k(t_0-t)+\mu\,C\eta\right] v(\gamma, \, \tau, \, \xi)=
  0.\\
\end{equation}
Obviously, the regular solutions $u$ and $v$ are proportional to
confluent hypergeometric functions \cite{Abramowitz}:
\begin{equation}\label{ugx}
    u(\gamma, \, \tau, \, \xi)=e^{\frac{i}{2}(\gamma-k)\xi}\,
    {_1F_1}\left(\frac12+i\tau, \, 1, \, -i\gamma\xi\right)
\end{equation}
and
\begin{equation}\label{vgx}
  \begin{array}{l}
    v(\gamma, \, \tau, \, \eta)=e^{-\frac{i}{2}(\gamma-k)\eta}\,
    {_1F_1}(\frac12+i(\tau-\tau_0), \, 1, \, i\gamma\eta)\hfill\\[3mm]
\hfill =e^{\frac{i}{2}(\gamma+k)\eta}\,
    {_1F_1}(\frac12+i(\tau_0-\tau), \, 1, \, -i\gamma\eta),\\
   \end{array}
\end{equation}
where
\begin{equation}\label{gtt0}
    \mu\,C=\frac{k^2}{2}-\frac{\gamma^2}{2},\quad \tau = \frac{k}{\gamma}\left(
    t+\frac{i}{2}\right), \quad \tau_0 =\frac{k}{\gamma}t_0.
\end{equation}
It should be noted that since the representation of the two- and
six-dimensional Coulomb Green's functions matrix elements involves
an integration over $\tau$ from $-\infty$ to $\infty$, in the
subsequent discussion we assume that $\tau$ is real. With this
assumption it is readily seen that the solutions (\ref{ugx}) and
(\ref{vgx}) coincide, except for normalization and the phase factors
$e^{-\frac{i}{2}k\xi}$ and $e^{\frac{i}{2}k\eta}$, with parabolic
Coulomb Sturmians treated in Ref. \cite{Gasaneo3}. In this case
$\gamma$ plays the role of the momentum and
$\mathcal{E}=\frac{\gamma^2}{2}$ is the energy. From this we
conclude that for $\gamma>0$ the solutions (\ref{ugx}) and
(\ref{vgx}) correspond to the continuous spectrum of $\mathcal{E}$.

It is readily verified that the solutions $u(\gamma, \, \tau, \,
\xi)$ and $v(\gamma, \, \tau, \, \eta)$ are expressed by basis set
(\ref{S1}) expansions
\begin{equation}\label{ugexp}
    u(\gamma,\,\tau,\,
    \xi)=\frac{2\sqrt{2b}}{2b-i(\gamma-k)}
    \left(\frac{2b-i(\gamma-k)}{2b+i(\gamma+k)}\right)^{i\tau+\frac12}
    \sum\limits_{n=0}^{\infty}\theta^n\,p_n\left(\tau;\;
    \zeta\right)\,\psi_n(\xi),
\end{equation}
\begin{equation}\label{vgexp}
    v(\gamma,\,\tau,\,
    \eta)=\frac{2\sqrt{2b}}{2b-i(\gamma+k)}
    \left(\frac{2b-i(\gamma+k)}{2b+i(\gamma-k)}\right)^{i(\tau_0-\tau)+\frac12}
    \sum\limits_{n=0}^{\infty}\lambda^{-n}\,p_n\left(\tau_0-\tau;\;
    \zeta\right)\,\psi_n(\eta),
\end{equation}
where
\begin{equation}\label{tlz}
     \theta=\frac{2b+i(\gamma-k)}{2b-i(\gamma-k)}, \qquad
     \lambda=\frac{2b-i(\gamma+k)}{2b+i(\gamma+k)}, \qquad
     \zeta=\frac{\lambda}{\theta}.
\end{equation}
The expansion (\ref{ugexp}) and (\ref{vgexp}) coefficients contain
the polynomials \cite{previous}
\begin{equation}\label{pnn}
 p_n\left(\tau;\;\zeta\right)=\frac{(-1)^n}{n!}\frac{\Gamma\left(n+\frac12
 -i\tau \right)}{\Gamma\left(\frac12-i\tau\right)}\;
    {_2F_1}{\left(-n,\,\frac12+i\tau;\;-n+\frac12+i\tau;\;\zeta\right)}.
\end{equation}

The basis set (\ref{S1}) representation of the equation (\ref{SYS1})
is the three-term recursion relation \cite{previous}
\begin{equation}\label{TRR}
    a_n\,y_{n-1}+b_n\,y_n+d_{n+1}\,y_{n+1}=0,
    \quad n \ge 1
\end{equation}
where
\begin{equation}\label{abd}
 \begin{array}{c}
  b_n=(b+\frac{\mu C}{2b}+ik)+2(b+\frac{\mu C}{2b})n+2kt,\\[2mm]
  a_n = (b-\frac{\mu C}{2b}-ik)n,\quad d_n = (b-\frac{\mu C}{2b}+ik)n.\\
 \end{array}
\end{equation}
The functions
\begin{equation}\label{sn}
    s_n(t;\; \mu C)=\theta^n\,p_n(\tau; \; \zeta)
\end{equation}
are the ``regular'' solutions of (\ref{TRR}) with the initial
conditions: $s_0 \equiv 1$, $s_{-1} \equiv 0$. The polynomials $p_n$
(\ref{pnn}) of degree $n$ in $\tau$ are orthogonal with respect to
the weight function \cite{previous}
\begin{equation}\label{rho}
    \rho(\tau; \zeta) = \frac{\Gamma\left(\frac12-i\tau\right)
    \Gamma\left(\frac12+i\tau\right)}{2\pi
    i}\,(-\zeta)^{i\tau+\frac12},
\end{equation}
where it is considered that $\left|\arg(-\zeta)\right|<\pi$. The
corresponding orthogonality relation reads
\begin{equation}\label{Orth1}
    \frac{i}{\zeta^m}\left(\frac{\zeta-1}{\zeta}\right)
    \int\limits_{-\infty}^{\infty}d \tau\,
    \rho(\tau;\; \zeta)\,p_n\left(\tau;\; \zeta\right)
    \,p_{m}\left(\tau;\; \zeta\right)=\delta_{n\,m}.
\end{equation}

\subsection{The discrete spectrum} For $t_0 < 0$ the
eigenfunctions of (\ref{SYS1}) corresponding to the discrete
spectrum: $\mathcal{E}^{(\ell)}=\frac{\gamma_{\ell}^2}{2}$,
$\gamma_{\ell}=i\kappa_{\ell}$, $\kappa_{\ell}=-\frac{k t_0}{\ell}$,
$\ell=1, 2, \,\ldots,\,\infty$, are \cite{Landau}
\begin{equation}\label{fl1l2}
    f_{\ell,\, m}(\xi,\, \eta)=u_{\ell,\, m}(\xi)\,v_{\ell,\,
    \ell-m-1}(\eta), \qquad m = 0, \, 1, \ldots\, , \, \ell-1,
\end{equation}
where
\begin{equation}\label{ulm}
    u_{\ell,\,m}(\xi)=e^{-\frac{i}{2}k \xi}e^{-\frac{\kappa_{\ell}
    \xi}{2}}\,
    {_1F_1}(-m, \, 1; \; \kappa_{\ell}\xi)=
    e^{-\frac{i}{2}k \xi}e^{-\frac{\kappa_{\ell}
    \xi}{2}}\,
    L_{m}(\kappa_{\ell}\xi)
\end{equation}
and
\begin{equation}\label{vlm}
    v_{\ell,\,m}(\eta)=e^{\frac{i}{2}k \eta}e^{-\frac{\kappa_{\ell}
    \eta}{2}}\,
    {_1F_1}(-m, \, 1; \; \kappa_{\ell}\eta)=
e^{\frac{i}{2}k \eta}e^{-\frac{\kappa_{\ell}
    \eta}{2}}\,
    L_{m}(\kappa_{\ell}\eta).
\end{equation}
The solutions $f_{\ell,\,m}$ meet the orthogonality relation
\begin{equation}\label{Orthf}
    \frac{\kappa_{\ell}^3}{2\ell}
    \int \limits _{0}^{\infty}\int \limits _{0}^{\infty}
    (\xi + \eta) d\xi d\eta\,f_{\ell,\, m}(\xi,\,
    \eta)\,
    \left[f_{{\ell\,}',\, m'}(\xi,\, \eta)\right]^{*}
    =\delta_{\ell,\, {\ell\,'}}\,\delta_{m,\, m'}.
\end{equation}

It is readily verified that the expansions of $u_{\ell,\,m}(\xi)$
and $v_{\ell,\,m}(\eta)$ in a basis function (\ref{S1}) series are
\begin{equation}\label{ulme}
    u_{\ell, \, m}(\xi)=\sum \limits_{n=0}^{\infty}
    \mathcal{S}^{(\ell,\, m)}_n\,\psi_n(\xi)
\end{equation}
and
\begin{equation}\label{vlme}
    v_{\ell, \, m}(\eta)=\sum \limits_{n=0}^{\infty}
    \left[\mathcal{S}^{(\ell,\, m)}_n\right]^*\psi_n(\eta),
\end{equation}
where the coefficients are given by
\begin{equation}\label{slm}
 \begin{array}{l}
     \mathcal{S}^{(\ell,\, m)}_n=2\sqrt{2b}\,(-)^n\,\frac{(m+1)_{n}}{n!}
    \frac{(2b-\kappa_{\ell}-i k)^n
    (2b-\kappa_{\ell}+i k)^{m}}
    {(2b+\kappa_{\ell}+i k)^{n+m+1}} \hfill\\[2mm]
\hfill    \times {_2F_1}\left(-n,\, -m, \, -n-m; \;
    \frac{(2b+\kappa_{\ell})^2+k^2}
    {(2b-\kappa_{\ell})^2+k^2}\right).\\
  \end{array}
\end{equation}

\subsection*{One-dimensional completeness relations}
The eigensolution $u(\gamma,\, \tau, \, \xi)$ of (\ref{SYS1})
corresponding to the continuous spectrum ($\gamma>0$) for large
$\xi$ behaves as
\begin{equation}\label{asymu}
    u(\gamma,\, \tau, \, \xi) \mathop{\sim}\limits_{\xi \rightarrow
    \infty}\,
    \frac{e^{-\frac{i}{2}k \xi}\,e^{\frac{\pi \tau}{2}}}{\left|\Gamma(\frac12+i
    \tau)\right|}\,
    \frac{2i}{\sqrt{\gamma \xi}}\, \sin\left(\frac{\gamma \xi}{2}-\tau
    \ln(\gamma \xi)+\frac{\pi}{4}+ \sigma
    \right),
\end{equation}
where $\sigma=\arg\Gamma\left(\frac12+i\tau\right)$. Therefore (see
e. g. \cite{Michel}),
\begin{equation}\label{I1dg}
    \frac{\gamma}{4}\left|\Gamma\left(\frac12+i
    \tau\right)\right|^2\, e^{-\pi \tau}\,\int\limits_{0}^{\infty}\xi\,d\xi\,
     u(\gamma, \, \tau, \,
    \xi)\,\left[u(\gamma', \, \tau, \, \xi)
    \right]^{*}=\pi\delta(\gamma-\gamma'),
\end{equation}
and for $\tau >0$
\begin{equation}\label{I1udx}
   \xi\int\limits_{0}^{\infty}\gamma\,d\gamma\,\frac{\left|\Gamma\left(\frac12+i
    \tau\right)\right|^2}{4\pi}\, e^{-\pi \tau}\,
     u(\gamma, \, \tau, \, \xi)\,\left[u(\gamma, \, \tau, \, \xi')
    \right]^{*}=\delta(\xi-\xi').
\end{equation}

On the other hand, if the functions $u(\gamma,\, \tau, \, \xi)$ are
regarded as charge Sturmians \cite{Gasaneo3}, i. e. the parameter
$\tau$ is considered as eigenvalue of the problem, whereas the
momentum $\gamma$ remains constant, the corresponding orthogonality
and completeness relations are given by \cite{Gasaneo3}
\begin{equation}\label{I2dt}
\gamma\,\left|\Gamma\left(\frac12+i\tau\right)\right|^2\,e^{-\pi
\tau} \int \limits_{0}^{\infty}d\xi\,
    u(\gamma, \, \tau, \, \xi)\left[u(\gamma, \, \tau', \,
    \xi)\right]^*=2\pi \delta(\tau-\tau')
\end{equation}
and
\begin{equation}\label{I2udx}
 \gamma\,\int \limits_{-\infty}^{\infty}d\tau\,
\frac{\left|\Gamma\left(\frac12+i\tau\right)\right|^2}{2\pi}\,e^{-\pi
\tau}\,u(\gamma, \, \tau, \, \xi)\left[u(\gamma, \, \tau, \,
    \xi')\right]^*=\delta(\xi-\xi').
\end{equation}
Taking matrix elements of the completeness relation (\ref{I2udx}) we
find
\begin{equation}\label{DAI2}
    \frac{8\,b\,\gamma\, \theta^{n-m}}
    {\sqrt{\left[4b^2+(\gamma+k)^2\right]\left[4b^2+(\gamma-k)^2\right]}}\,
    \int \limits_{-\infty}^{\infty}d\tau\,
\frac{\left|\Gamma\left(\frac12+i\tau\right)\right|^2}{2\pi}\,(-\zeta)^{i\tau}
p_n\left(\tau;\;\zeta\right) \left[ p_n\left(\tau;\;
\zeta\right)\right]^*=\delta_{n,\,
m}. 
\end{equation}
It may be noted that (\ref{DAI2}) is closely related to
(\ref{Orth1}). To see this, let $\gamma>0$. Then, it follows from
the definitions (\ref{tlz}) that
\begin{equation}\label{fc1}
    \frac{(\zeta-1)}{\zeta}(-\zeta)^{\frac12}=\frac{8\,b\,\gamma}
    {\sqrt{\left[4b^2+(\gamma+k)^2\right]\left[4b^2+(\gamma-k)^2\right]}}.
\end{equation}
Further, the regular solution $\theta^n p_n\left(\tau;\;
\zeta\right)$ of the three-term recursion relation (\ref{TRR}) is an
even function of $\gamma$, since the coefficients $a_n$, $b_n$ and
$d_n$ depend on $\gamma$ only through $\mu
C=\frac{1}{2}(k^2-\gamma^2)$. Thus, replacing $\gamma$ by $-\gamma$,
and hence $\tau$ by $-\tau$ ($\theta \rightarrow \lambda$ and
$\lambda \rightarrow \theta$) and $\zeta$ by $\zeta^{-1}$ in Eq.
(\ref{pnn}) gives
\begin{equation}\label{pgp}
\theta^n\,p_n\left(\tau;\; \zeta\right)=\lambda^n
\frac{(-1)^n}{n!}\frac{\Gamma\left(n+\frac12
 +i\tau \right)}{\Gamma\left(\frac12+i\tau\right)}\;
    {_2F_1}{\left(-n,\,\frac12-i\tau;\;-n+\frac12-i\tau;\;\zeta^{-1}\right)}.
\end{equation}
Comparing Eqs. (\ref{pnn}) and (\ref{pgp}) then yields the relation
\begin{equation}\label{pa1}
\theta^n\,p_n\left(\tau;\; \zeta\right)=
\lambda^{n}\left[p_n\left(\tau;\; \zeta\right)\right]^*,
\end{equation}
and hence
\begin{equation}\label{pa2}
 \left[p_n\left(\tau;\; \zeta\right)\right]^*
 =\zeta^{-n}\,p_n\left(\tau;\; \zeta\right).
\end{equation}
From the argument above, we conclude that for $\gamma>0$ the
orthogonality relation (\ref{Orth1}) reduces to (\ref{DAI2}).

\subsection*{The two-dimensional completeness relation}
It follows from the relations (\ref{I1dg}) and (\ref{I2dt}) and
analogous relations for $v(\gamma, \, \tau, \, \xi)$ that the
two-dimensional orthogonality relation for $\gamma>0$ is given by
\begin{equation}\label{I12dgdt}
 \begin{array}{c}
    e^{-\pi\tau_0}\,\frac{\gamma^2}{4}\,
    \frac{\left|\Gamma\left(\frac12+i\tau\right)\right|^2}{2\pi}
    \frac{\left|\Gamma\left(\frac12+i(\tau_0-\tau)\right)\right|^2}{2\pi}\,
    \int \limits_{0}^{\infty}\int
    \limits_{0}^{\infty}(\xi+\eta)\,d\xi d\eta \left\{f(\gamma, \, \tau, \, \xi, \,
    \eta) \right. \qquad \qquad \qquad  \\[3mm]
    \qquad \qquad \qquad \left. \times \left[f(\gamma', \, \tau', \, \xi, \,
    \eta)\right]^*\right\}=\delta(\gamma-\gamma')\,\delta(\tau-\tau').\\
 \end{array}
\end{equation}
In turn, in the case $t_0>0$ (where there are no bound states) it
would appear reasonable that the two-dimensional completeness
relation would be given by
\begin{equation}\label{CR0}
 \begin{array}{l}
(\xi+\eta) \left\{ \alpha \int \limits_{0}^{\infty}d\gamma\,
\gamma^2 e^{-\pi\tau_0}
  \int \limits_{-\infty}^{\infty}d\tau
\frac{\left|\Gamma\left(\frac12+i\tau \right) \right|^2}{2\pi}
\frac{\left|\Gamma\left(\frac12+i(\tau_0-\tau) \right)
\right|^2}{2\pi} \right.\hfill \\[3mm]
\hfill \left.\times f(\gamma, \, \tau, \, \xi, \, \eta)
\left[f(\gamma, \, \tau, \, \xi', \, \eta')\right]^* \vphantom{\int
\limits_{-\infty}^{\infty}} \right\}
    = \delta(\xi-\xi')\,\delta(\eta-\eta').\\
 \end{array}
\end{equation}
The integration over $\tau$ in (\ref{CR0}) is performed on the
assumption that $\tau$ is independent of $\gamma$. To test this
hypothesis and determine the normalizing factor $\alpha$, we carried
out the following numerical experiments. First with some parameters
$t_0>0$, $k>0$ and $b$ we calculate the matrix elements $A_{n_1, \,
n_2;\; m_1, \, m_2}$ for the expression in the figure braces on the
right-hand side of (\ref{CR0}) in the basis (\ref{S2}):
\begin{equation}\label{Anm}
 \begin{array}{l}
 A_{n_1, \, n_2;\; m_1, \, m_2}=
\alpha \int \limits_{0}^{\infty}d\gamma\, \frac{64\,b^2
\,\gamma^2\,(-\zeta)^{i\tau_0}}{\left[4b^2+(\gamma+k)^2\right]
\left[4b^2+(\gamma-k)^2\right]}
\,\theta^{n_1-m_1}\lambda^{m_2-n_2}\hfill\\[3mm]
\times \int \limits _{-\infty}^{\infty} d \tau
\frac{\left|\Gamma\left(\frac12+i\tau \right) \right|^2}{2\pi}
\frac{\left|\Gamma\left(\frac12+i(\tau_0-\tau) \right)
\right|^2}{2\pi}\,p_{n_1}\left(\tau;\; \zeta\right)
p_{n_2}\left(\tau_0-\tau;\; \zeta\right)\left[p_{m_1}\left(\tau;\;
\zeta\right)
p_{m_2}\left(\tau_0-\tau;\; \zeta\right)\right]^*.\\
 \end{array}
\end{equation}
It should be noted that the value of $(-\zeta)^{i\tau_0}$ in this
formula is determined by the condition $\left|\arg(-\zeta)\right|<
\pi$. Then the resulting matrix ${\bf A}$ is multiplied by the
matrix
\begin{equation}\label{QQ}
{\bf Q}={\bf Q}_{\xi}\otimes{\bf I}_{\eta}+{\bf I}_{\xi}\otimes{\bf
Q}_{\eta}
\end{equation}
of the operator $(\xi+\eta)$. Finally, using the condition
\begin{equation}\label{QA2}
  {\bf Q}\,{\bf A}={\bf I}_{\xi}\otimes{\bf I}_{\eta}.
\end{equation}
we have obtained that $\alpha=\frac12$. Notice that the infinite
symmetric matrices ${\bf Q}_{\xi}$ and ${\bf Q}_{\eta}$ are
tridiagonal \cite{previous}, therefore the equations on the left
hand-side of the linear system (\ref{QA2}) each contain no more than
six terms.

For for $t_0<0$ the completeness relation (\ref{CR0}) transforms
into
\begin{equation}\label{CR1}
 \begin{array}{c}
(\xi+\eta) \left\{ \frac12\int \limits_{0}^{\infty}d\gamma\,
\gamma^2 e^{-\pi\tau_0}
  \int \limits_{-\infty}^{\infty}d\tau
\frac{\left|\Gamma\left(\frac12+i\tau \right) \right|^2}{2\pi}
\frac{\left|\Gamma\left(\frac12+i(\tau_0-\tau) \right)
\right|^2}{2\pi}\,f(\gamma, \, \tau, \, \xi, \, \eta)
\left[f(\gamma, \, \tau, \, \xi', \, \eta')\right]^* \right.\\[3mm]
\left.    \hfill+\sum \limits _{\ell=1}^{\infty}
    \frac{\kappa_{\ell}^3}{2\ell}\,\sum \limits _{m=0}^{\ell-1}
    f_{\ell,\, m,\, \ell-m-1}(\xi,\, \eta)
    \left[f_{\ell,\, m,\, \ell-m-1}(\xi',\, \eta')\right]^* \right\}
    = \delta(\xi-\xi')\,\delta(\eta-\eta').\\
 \end{array}
\end{equation}
In this case the matrix ${\bf A}$ with elements
\begin{equation}\label{Anm1}
 \begin{array}{l}
 A_{n_1, \, n_2;\; m_1, \, m_2}=
\int \limits_{0}^{\infty}d\gamma\, \frac{32\,b^2
\,\gamma^2\,(-\zeta)^{i\tau_0}}
{\left[4b^2+(\gamma+k)^2\right]\left[4b^2+(\gamma-k)^2\right]}
\,\theta^{n_1-m_1}\lambda^{m_2-n_2}\hfill\\[3mm]
\times \int \limits _{-\infty}^{\infty} d \tau
\frac{\left|\Gamma\left(\frac12+i\tau \right) \right|^2}{2\pi}
\frac{\left|\Gamma\left(\frac12+i(\tau_0-\tau) \right)
\right|^2}{2\pi}\,p_{n_1}\left(\tau;\; \zeta\right)
p_{n_2}\left(\tau_0-\tau;\; \zeta\right)\left[p_{m_1}\left(\tau;\;
\zeta\right)p_{m_2}\left(\tau_0-\tau;\; \zeta\right)\right]^*
\\[3mm]
    \hfill+\sum \limits _{\ell=1}^{\infty}
    \frac{\kappa_{\ell}^3}{2\ell}\,\sum \limits _{m=0}^{\ell-1}
    \mathcal{S}^{(\ell,\, m)}_{n_1}\,\mathcal{S}^{(\ell,\,\ell-m-1)}_{m_2}
    \left[\mathcal{S}^{(\ell,\,
    \ell-m-1)}_{n_2}\, \mathcal{S}^{(\ell,\,m)}_{m_1}\right]^*,\\
 \end{array}
\end{equation}
is also inverse to the matrix ${\bf Q}$ (\ref{QQ}). The expression
(\ref{Anm1}) can be rewritten, in view of (\ref{fc1}) and
(\ref{pa2}), as
\begin{equation}\label{Anm2}
 \begin{array}{l}
 A_{n_1, \, n_2;\; m_1, \, m_2}=
-\frac12 \int
\limits_{0}^{\infty}d\gamma\left\{\left(\frac{\zeta-1}{\zeta}
\right)^2\,\frac{\theta^{n_1+m_2}}{\lambda^{n_2+m_1}}\, \int \limits
_{-\infty}^{\infty} d \tau
\rho(\tau;\; \zeta)\,\rho(\tau_0-\tau;\; \zeta) \right.\\[3mm]
\hfill \left. \times p_{n_1}(\tau;\; \zeta)\,p_{n_2}(\tau_0-\tau;\;
\zeta)\, p_{m_1}(\tau;\; \zeta)\, p_{m_2}(\tau_0-\tau;\; \zeta)
\vphantom{\left(\frac{\zeta-1}{\zeta} \right)^2}\right\}
\\[3mm]
    \hfill+\sum \limits _{\ell=1}^{\infty}
    \frac{\kappa_{\ell}^3}{2\ell}\,\sum \limits _{m=0}^{\ell-1}
    \mathcal{S}^{(\ell,\, m)}_{n_1}\,\mathcal{S}^{(\ell,\,\ell-m-1)}_{m_2}
    \left[\mathcal{S}^{(\ell,\,
    \ell-m-1)}_{n_2}\, \mathcal{S}^{(\ell,\,m)}_{m_1}\right]^*.\\
 \end{array}
\end{equation}

\section{Contour integrals}
Notice that expressing the resolvent of the one-dimensional operator
$\left[ \hat{h}_{\xi}+2kt+\mu\,C\xi\right]$ requires two linearly
independent solutions of (\ref{SYS1}). Irregular solutions of
(\ref{SYS1}) are expressed in terms the confluent hypergeometric
function \cite{Abramowitz}
\begin{equation}\label{IS}
    w^{(\pm)}(\gamma,\, \tau, \, \xi)=e^{\frac{i}{2}(\pm
    \gamma-k)\xi}\,{U\left(\frac12\pm i\tau,\, 1;\; \mp\gamma \xi
    \right)}.
\end{equation}
The corresponding solutions of the  three-term recursion relation
(\ref{TRR}) are
\begin{equation}\label{cn}
 \begin{array}{c}
     c_n^{(+)}(t;\; \mu C)
    =\theta^{n+1}\,q^{(+)}_n\left(\tau;\; \zeta\right),\\[3mm]
    c_n^{(-)}(t;\; \mu C)
    =\lambda^{n+1}\,q^{(-)}_n\left(\tau;\; \zeta\right),\\
\end{array}
\end{equation}
where
\begin{equation}\label{qn}
 \begin{array}{c}
     q_n^{(+)}(\tau;\; \zeta)=(-)^n\,\frac{n!\,\Gamma\left(\frac12+i\tau\right)}
    {\Gamma\left(n+\frac32+i\tau\right)}\;
    {_2F_1}\left(\frac12+i\tau,\, n+1;\; n+\frac32+i\tau;\;
    \zeta^{-1}\right),\\[3mm]
    q_n^{(-)}(\tau; \; \zeta)=(-)^n\,\frac{n!\,\Gamma\left(\frac12-i\tau\right)}
    {\Gamma\left(n+\frac32-i\tau\right)}\;
    {_2F_1}\left(\frac12-i\tau,\, n+1;\; n+\frac32-i\tau;\;
    \zeta\right).\\
\end{array}
\end{equation}
In particular, the functions
\begin{equation}\label{wpexp}
 \begin{array}{l}
    \widetilde{w}^{(\pm)}(\gamma,\,\tau,\,
    \xi)=\frac{2i \sqrt{2b}}{2b-i(\gamma-k)}
    \left(-\frac{2b+i(\gamma-k)}{2b-i(\gamma+k)}\right)^{i\tau+\frac12}
    \hfill \\[3mm]
\hfill    \times    \frac{e^{-\pi
\tau}}{\theta\,\Gamma\left(\frac12\pm i\tau \right)}
    \sum\limits_{n=0}^{\infty}c_n^{(\pm)}(t;\; \mu
    C)\,\psi_n(\xi)\\
 \end{array}
\end{equation}
tend to $w^{(\pm)}(\gamma,\,\tau,\,\xi)$ as $\xi \rightarrow
\infty$.

The matrix elements of the resolvent of $\left[
\hat{h}_{\xi}+2kt+\mu\,C\xi\right]$ can be written in the form
\cite{previous}
\begin{equation}\label{gnm_p}
    g^{(+)}_{n,\, m}(t;\; \mu
    C)=\frac{i}{2\gamma}\left(\frac{\zeta-1}{\zeta}\right)\frac{\theta^{n-m}}{\zeta^m}\,
    p_{n_{<}}\left(\tau;\; \zeta\right)\,q^{(+)}_{n_{>}}(\tau;\; \zeta)
\end{equation}
and
\begin{equation}\label{gnm_m}
    g^{(-)}_{n,\, m}(t;\; \mu
    C)=\frac{i}{2\gamma}\left(\frac{\zeta-1}{\zeta}\right)\frac{\theta^{n-m}}{\zeta^m}\,
    p_{n_{<}}\left(\tau;\; \zeta\right)\,\zeta^{n_{>}+1}\,q^{(-)}_{n_{>}}(\tau;\; \zeta),
\end{equation}
where $n_{>}$ and $n_{<}$ are the greater and lesser of $n$ and $m$.
Notice that $c_n^{(+)}$ $\left( c_n^{(-)}\right)$ are defined in the
upper (lower) half of the complex $\gamma$-plane where $|\zeta|\geq
1$ $\left( |\zeta| \leq 1\right)$. To analytically continue
$c_n^{(+)}$ onto the lower half of the $\gamma$-plane the relation
\cite{previous}
\begin{equation}\label{cpm}
    c_n^{(+)}(t;\; \mu C)=c_n^{(-)}(t;\; \mu C)+2\pi i \rho(\tau; \;
    \zeta)\,\theta^{n+1}p_n\left(\tau;\; \zeta\right)
\end{equation}
can be used.

In \cite{previous} we obtained the basis set (\ref{S2})
representation of the resolvent for the two-dimensional operator
$\left[ \hat{h}_{\xi}+2kt+\mu\,C\xi\right]+\left[
\hat{h}_{\eta}+2k(t_0-t)+\mu\,C\eta\right]$. In particular, the
matrix elements of the two-dimensional Green's function can be
expressed as the convolution integral
\begin{equation}\label{G3}
 \begin{array}{l}
G^{(\pm)}_{n_1,\,n_2;\; m_1,\,m_2}(t_0;\; \mu C)
=i\left(\frac{\zeta-1}{\zeta}\right)\,\frac{\lambda^{m_2-n_2}}{\zeta^{m_2}}
\int \limits _{-\infty}^{\infty} d\tau\,
    \rho(\tau_0-\tau;\; \zeta)\\[3mm]
\hfill \times g^{(\pm)}_{n_1,\, m_1}(t;\; \mu C)\,
    p_{n_2}(\tau_0-\tau;\; \zeta)\,p_{m_2}(\tau_0-\tau;\; \zeta).
 \end{array}
\end{equation}
Notice that in this case only the regular solutions of (\ref{SYS2})
discrete analogues $\lambda^{-n}p_n(\tau_0-\tau;\; \zeta)$ are used.

Let us consider the integral
\begin{equation}\label{mgpg}
   \mathcal{I}_1=\frac{1}{2\pi i}\int \limits _{-\infty}^{\infty}\gamma\,d\gamma\,
   G^{(+)}_{n_1,\,n_2;\; m_1,\,m_2}(t_0;\; \mu C).
\end{equation}
Notice that by replacing $\gamma \rightarrow -\gamma$ (and hence
$\theta \rightarrow \lambda$, $\lambda \rightarrow \theta$, $\zeta
\rightarrow 1/\zeta$, $\tau \rightarrow -\tau$ and $\tau_0
\rightarrow -\tau_0$) in Eq. (\ref{G3}) $G^{(+)}_{n_1,\,n_2;\;
m_1,\,m_2}(t_0;\; \mu C)$ is transformed to $G^{(-)}_{n_1,\,n_2;\;
m_1,\,m_2}(t_0;\; \mu C)$. Thus, for the integral $\mathcal{I}_1$ we
obtain
\begin{equation}\label{mgpg2}
\mathcal{I}_1=\frac{1}{2\pi i}\int \limits
_{0}^{\infty}\gamma\,d\gamma \left\{G^{(+)}_{n_1,\,n_2;\;
m_1,\,m_2}(t_0;\; \mu C)-G^{(-)}_{n_1,\,n_2;\; m_1,\,m_2}(t_0;\; \mu
C) \right\}.
\end{equation}
Inserting Eqs. (\ref{G3}), (\ref{gnm_p}) and (\ref{gnm_m}) into Eq.
(\ref{mgpg2}), we find, in view of (\ref{cpm}), that $\mathcal{I}_1$
coincides with the integral on the right hand-side of (\ref{Anm2}).

Now, we consider the integral
\begin{equation}\label{Cont}
    \frac{1}{2\pi i}\int \limits _{\mathcal{C}}\,
    d\mathcal{E}\,{\bf G}^{(+)}\left(t_0;\; \frac{k^2}{2}
    -\mathcal{E}\right),
\end{equation}
taken along the contour in the complex $\mathcal{E}$-plane shown in
Fig.~1. The contour $\mathcal{C}$ passes in a negative direction
(clockwise) round all the points
$\mathcal{E}^{(\ell)}=-\frac{\kappa_{\ell}^2}{2}$ (filled circles in
Fig.~1 which accumulate at the origin) and the cut along the right
half of the real axis and is closed at infinity (see e. g.
\cite{Shakeshaft}). The corresponding matrix element of the integral
along the two sides of the cut is equal to $\mathcal{I}_1$
(\ref{mgpg}) (this is circumstantial evidence that the normalizing
factor $\alpha$ in the completeness relation (\ref{CR0}) is equal to
$\frac12$). On the other hand, the integration along a contour
enclosing $\mathcal{E}^{(\ell)}$ reduces to $(-1)$ times the double
sum of the residues of the integrand at the points
$\tau^{(m)}=i\left(m+\frac12\right)$, $m=0, \, 1,\, \ldots$ and
${\mathcal{E}^{(\ell)}=-\frac{(kt_0)^2}{2\ell^2}}$, $\ell=1,\, 2,\,
\ldots\,$, which are the poles of the gamma functions
$\Gamma\left(\frac12+i\tau \right)$ and
$\Gamma\left(\frac12+i(\tau_0-\tau) \right)=\Gamma\left(m+1+\frac{k
t_0}{\sqrt{-2\mathcal{E}^{(\ell)}}} \right)$, respectively. It is
readily shown that the matrix element of this part of the integral
(\ref{Cont}) coincides with the double sum in (\ref{Anm2}). Thus,
the integral (\ref{Cont}) is equal to the matrix ${\bf A}$. The
contour $\mathcal{C}$ can be deformed, for instance, into a straight
line parallel to the real axis. The resulting path $\mathcal{C}_1$
shown in Fig.~2 runs above the cut and the bound-state poles of
${\bf G}^{(+)}\left(t_0;\; \frac{k^2}{2} -\mathcal{E}\right)$. Then,
to make the integral amenable to numerical integration, rotate
$\mathcal{C}_1$ about some point $\mathcal{E}_0$ on the right half
of the real axis through a negative angle $\varphi$
\cite{Shakeshaft}; see the contour $\mathcal{C}_2$ in Fig.~2. The
part of $\mathcal{C}_2$ on the unphysical sheet is depicted by the
dashed line. Thus, we obtain the following representation of the
matrix ${\bf A}$ (which is inverse to the matrix ${\bf Q}$
(\ref{QQ})):
\begin{equation}\label{IQ}
    {\bf A}=
    \frac{1}{2\pi i}\int \limits _{\mathcal{C}_2}\,
    d\mathcal{E}\,{\bf G}^{(+)}\left(t_0;\; \frac{k^2}{2}
    -\mathcal{E}\right)={\bf Q}^{-1}.
\end{equation}

\section{Six-dimensional Green's function matrix}
Using the relation (\ref{IQ}) we can rewrite the expression
(\ref{G6}) for the six-dimensional Coulomb Green' function matrix as
the contour integral
\begin{equation}\label{G62}
\underline{\mathfrak{G}}= \frac{\aleph}{\mu_{23}\mu_{13}} \int
\limits_{\mathcal{C}_2}d\mathcal{E}_1 \int\limits_{\mathcal{C}_2}
d\mathcal{E}_2\,{\bf G}_1\left(t_{23};\;
\frac{k_{23}^2}{2}-\mathcal{E}_1\right)\otimes
    {\bf G}_2\left(t_{13};\; \frac{k_{13}^2}{2}-\mathcal{E}_2\right)
 \otimes{\bf G}_3\left(t_{12};\; \frac{k_{12}^2}{2}-\mathcal{E}_3\right), 
\end{equation}
where $\mathcal{E}_j=\frac{k_{ls}^2}{2}-\mu_{ls}C_j$ and
$C_1+C_2+C_3=0$. This also allows us to determine the normalizing
factor $\aleph$. Indeed, it follows from (\ref{IQ}) that the third
term inside the figure brackets on the right-hand side of
(\ref{hG3}) is proportional to the unit matrix:
\begin{equation}\label{ThirdT}
 \begin{array}{c}
\left[{\bf Q}_1\int \limits_{\mathcal{C}_2} d\mathcal{E}_1\,{\bf
G}_1\left(t_{23};\; \frac{k_{23}^2}{2}-\mathcal{E}_1\right)\right]
\otimes\left[{\bf Q}_2\int \limits_{\mathcal{C}_2}
d\mathcal{E}_2\,{\bf
G}_2\left(t_{13};\;\frac{k_{13}^2}{2}-\mathcal{E}_2\right)
 \right]\otimes{\bf I}_3\\[3mm]
=(2\pi i)^2\,{\bf
I}_1\otimes{\bf I}_2\otimes{\bf I}_3.\\
\end{array}
\end{equation}
Consider the first two terms in the figure braces in (\ref{hG3}).
For the energy $\mathcal{E}_3=\frac{k_{12}^2}{2}+\mu_{12}(C_1+C_2)$
we have
\begin{equation}\label{E3}
\mathcal{E}_3= \frac{k_{12}^2}{2}+
\frac{\mu_{12}}{\mu_{23}}\left(\frac{k_{23}^2}{2}-\mathcal{E}_1\right)
+\frac{\mu_{12}}{\mu_{13}}\left(\frac{k_{13}^2}{2}-\mathcal{E}_2\right).
\end{equation}
On the other hand, on the contours $\mathcal{C}_2$ the energy
variables $\mathcal{E}_j$, $j=1,\, 2$ are given by
\begin{equation}\label{Ej}
    \mathcal{E}_j=\mathcal{E}_{0j}+E_j e^{i\varphi},
\end{equation}
where $\varphi<0$, $\mathcal{E}_{0j}$ is an arbitrary positive
parameter, $E_j$ is real and runs from $-\infty$ to $\infty$. Hence,
for the energy $\mathcal{E}_3$ (\ref{E3}) we obtain
\begin{equation}\label{E31}
\mathcal{E}_3=\left[\frac{k_{12}^2}{2}
+\frac{\mu_{12}}{\mu_{23}}\left(\frac{k_{23}^2}{2}-\mathcal{E}_{01}\right)
+\frac{\mu_{12}}{\mu_{13}}\left(\frac{k_{13}^2}{2}-\mathcal{E}_{02}\right)\right]
+\left(-\frac{\mu_{12}}{\mu_{23}}\,E_1-
\frac{\mu_{12}}{\mu_{13}}\,E_2\right)e^{i\varphi}.
\end{equation}
Thus, $\mathcal{E}_3$ can be expressed in the form
\begin{equation}\label{E32}
\mathcal{E}_3=\mathcal{E}_{03}+E_3\,e^{i\varphi},
\end{equation}
where $\mathcal{E}_{03}$ and $E_3$ denote the term in the square
braces and the real factor in front of the exponent in (\ref{E31}),
respectively. Since $\mathcal{E}_{03}$ should be positive, the
positive parameters $\mathcal{E}_{01}$ and $\mathcal{E}_{02}$ have
to satisfy the constraint
\begin{equation}\label{E01E02C}
\frac{\mu_{12}}{\mu_{23}}\,\mathcal{E}_{01}+\frac{\mu_{12}}{\mu_{13}}\,\mathcal{E}_{02}<
\frac{k_{12}^2}{2} +\frac{\mu_{12}}{\mu_{23}}\frac{k_{23}^2}{2}
+\frac{\mu_{12}}{\mu_{13}}\frac{k_{13}^2}{2}.
\end{equation}

Now, we consider the integral
\begin{equation}\label{I2}
    \mathcal{I}_2=\int
\limits_{\mathcal{C}_2} d\mathcal{E}_1 {\bf G}_3\left(t_{12};\;
\frac{k_{12}^2}{2}-\mathcal{E}_3\right).
\end{equation}
With fixed $\mathcal{E}_2$, in view (\ref{E31}), (\ref{E32}) and
(\ref{IQ}), we see that
\begin{equation}\label{I22}
\begin{array}{c}
    \mathcal{I}_2= e^{i\varphi}\int
\limits_{-\infty}^{\infty} d E_1 {\bf G}_3\left(t_{12};\;
\frac{k_{12}^2}{2}-\mathcal{E}_3\right)=-\frac{\mu_{23}}{\mu_{12}}\,
e^{i\varphi}\int \limits_{-\infty}^{\infty} d E_3 {\bf
G}_3\left(t_{12};\; \frac{k_{12}^2}{2}-\mathcal{E}_3\right)\\[3mm]
=-\frac{\mu_{23}}{\mu_{12}}\int \limits_{\mathcal{C}_2}
d\mathcal{E}_3 {\bf G}_3\left(t_{12};\;
\frac{k_{12}^2}{2}-\mathcal{E}_3\right)=-2\pi i\,
\frac{\mu_{23}}{\mu_{12}}\,{\bf Q}_3^{-1}.
 \end{array}
\end{equation}
Similarly, we obtain
\begin{equation}\label{I23}
\int \limits_{\mathcal{C}_2} d\mathcal{E}_2 {\bf G}_3\left(t_{12};\;
\frac{k_{12}^2}{2}-\mathcal{E}_3\right)=-2\pi
i\,\frac{\mu_{13}}{\mu_{12}}\,{\bf Q}_3^{-1}.
\end{equation}
Inserting (\ref{I22}), (\ref{I23}) and (\ref{ThirdT}) into
(\ref{hG3}) then yields
\begin{equation}\label{hG6}
    \underline{\mathfrak{h}}\;\underline{\mathfrak{G}}= \aleph\,4
    \pi^2\,{\bf
I}_1\otimes{\bf I}_2\otimes{\bf I}_3.
\end{equation}
Thus, from (\ref{hG6}) we conclude that
\begin{equation}\label{NF}
    \aleph=\frac{1}{4\pi^2}.
\end{equation}

\section{Conclusion}
The Sturmian basis-set representation of the resolvent for the
asymptotic three-body Coulomb wave operator is obtained, which can
be used in the discrete analog of the Lippmann-Schwinger equation
for the three-body continuum wave function. The six-dimensional
Green's function matrix is expressed as a convolution integral over
separation constants. The integrand of this contour integral
involves Green's function matrices corresponding to the
two-dimensional operators which are constituents of the full
six-dimensional wave operator. The completeness relation of the
eigenfunctions of these two-dimensional operators is used to define
the appropriate pathes of integration of the convolution integral.

\end{document}